\documentclass[aps,pra,superscriptaddress,twocolumn]{revtex4-1} 
\usepackage{amsmath, bm}
\usepackage{amssymb}
\usepackage{bm}
\usepackage{epsfig}
\usepackage{graphicx}
\usepackage{color}
\usepackage[normalem]{ulem}
\usepackage[makeroom]{cancel}

\newcommand{\lan}{\langle}
\newcommand{\ran}{\rangle}

\DeclareMathOperator{\sech}{sech}

\begin{document}

\title{Rabi-coupling driven motion of a soliton in a Bose-Einstein condensate}

\author{Sh. Mardonov}
\affiliation{Department of Physical Chemistry, The University of the Basque Country UPV/EHU,
48080 Bilbao, Spain}
\affiliation{The Samarkand State University, 140104 Samarkand, Uzbekistan}
\author{M. Modugno}
\affiliation{IKERBASQUE Basque Foundation for Science, Bilbao, Spain}
\affiliation{Department of Theoretical Physics and History of Science, University of the
Basque Country UPV/EHU, 48080 Bilbao, Spain}
\author{E. Ya. Sherman}
\affiliation{Department of Physical Chemistry, The University of the Basque Country UPV/EHU,
48080 Bilbao, Spain}
\affiliation{IKERBASQUE Basque Foundation for Science, Bilbao, Spain}
\author{B.A. Malomed}
\affiliation{Department of Physical Electronics, 
School of Electrical Engineering and Center for Light-Matter Interaction, Tel Aviv University, Tel Aviv 69978, Israel}

\begin{abstract}
We study {the} motion of {a self-attractive} Bose-Einstein condensate {with pseudo-spin $1/2$} 
driven by a synthetic Rabi (Zeeman-like) field. This field {triggers} the pseudo-spin dynamics
resulting in {a} density redistribution between its components and, as a consequence,
in changes of the overall density distribution. In the presence of an additional external potential, the latter produces
a net force acting on the condensate and activates its displacement. As an example, 
here we consider the case of a one-dimensional condensate in a random potential. 
\end{abstract}

\date{\today}

\maketitle

\section{Introduction}

The dynamics of self-interacting quantum matter in a random potential is a topic 
of a great significance \cite{Flach2009,Aleiner2010,Lucioni2011}. Adding a spin degree of freedom and spin-orbit 
coupling (SOC) considerably extends
the variety of patterns featured by these settings. The resulting coupled spin and mass
density motion is one of the most interesting manifestations of the underlying SOC \cite{Leurs2008}, 
where the particle spin is directly coupled to its momentum,
and the spin evolution naturally drives changes in the particle position,
both for solid-state \cite{Schliemann2005,Zawadzki2005,Winkler2007} and cold atom realizations 
alike \cite{zhaih2012,spielman2013}. The same mechanism may determine the 
motion of matter-wave solitons \cite{Beijing}.

Taking {a self-attractive} two-component Bose-Einstein condensate (BEC),
which constitutes a pseudo-spin $1/2$ system, as an example, 
we show here that such a mutual dependence of pseudo-spin and position can occur 
even without SOC provided that the BEC symmetry with respect to the spin rotations
is lifted by particle-particle interactions. 
This effect occurs in generic situation when the inter- and {intra-species} couplings are not equal, 
resulting in the non SU(2)-symmetric nonlinearity, as it is often 
{the case for} mean field interaction in binary BECs.  Then, the
pseudospin-dependent force driving the BEC may appear as a joint 
result of the Rabi (Zeeman) coupling acting on the atomic hyperfine
states, and, thus, affecting the BEC shape, and an external random potential
into which the BEC is loaded.

\section{Model and main parameters} We consider a quasi one-dimensional condensate
in the presence of a synthetic Rabi (Zeeman) field applied along the $x-$direction 
and of a spin-diagonal random potential $U(x).$ 
The two-component pseudospinor wave function, 
${\bm \psi}(\mathbf{x})\equiv\left[\psi_{1}(\mathbf{x}),\psi_{2}(\mathbf{x})\right]^{\rm T}$ 
(${\rm T}$ stands for transpose, ${\bf x}\equiv(x,t)$)
obeys {two} Gross-Pitaevskii equations {($\nu,\nu^{\prime}=1,2$)}
\begin{eqnarray}
i\partial_{t}{\psi_{\nu}}(\mathbf{x}) &=&{\left[-\frac{1}{2}\partial_{xx}+ U(x)\right]}{\psi_{\nu}}(\mathbf{x})- \label{eq:GPE}\\
&& \left(g|{\psi_{\nu}}(\mathbf{x})|^{2}+\tilde{g}|{\psi_{\nu^{\prime}}}(\mathbf{x})|^{2}\right)\psi_{\nu}(\mathbf{x})+
\frac{\Delta}{2}\psi_{\nu^{\prime}}(\mathbf{x}), \nonumber
\end{eqnarray}
where $\Delta$ is the Zeeman splitting and 
$g,\tilde{g}>0$ are the interatomic interaction constants. 
Units are chosen such that $\hbar=M=N=1$, where $M$ is the particle mass,
and $N$ is the norm. 
In the absence of {the random} potential, this model has been studied extensively 
in nonlinear optics of dual-core fibers (albeit in the time domain), where $\tilde{g}=0$, 
with $\Delta$ corresponding to the 
coupling between the fibers \cite{Uzunov1995,Malomed2018} and
in Rabi-coupled BECs \cite{Williams2000,Sartori2015,sakaguchi-radio-2D1,sakaguchi-radio-2D2}, where both 
$g$ and $\tilde{g}$ present.
An implementation with small random variations of $\Delta(t)$ has been
{considered in} Ref. [\onlinecite{Mostofi1997}].

{Here we} describe the system evolution {by means of} the density
matrix ${\bm\rho}(\mathbf{x})\equiv\,{\bm\psi(\mathbf{x})}{\bm\psi^{\dagger}(\mathbf{x})}$
and obtain observables by corresponding tracing.
We characterize the condensate motion by the center of mass
position $X(t)$:
\begin{equation}
X(t)={\rm tr}\int_{-\infty}^{\infty}x{\bm\rho}({\mathbf x})dx,
\label{eq:Xt}
\end{equation}
and {the} spin components {$\sigma_{i}(t)$} (here {$i=x,y,z$}) as:
\begin{equation}
\sigma_{i}(t) = {\rm tr}\int_{-\infty}^{\infty}\hat{\sigma}_{i}{\bm\rho}({\mathbf x})dx,
\label{eq:sigmait}
\end{equation}
where $\hat{\sigma}_{i}$ are the Pauli matrices. {For a general description of the spin state 
we introduce its squared length: 
$P(t)=\sum_{i}\sigma_{i}^{2}(t).$
When the two spinor components  are linearly dependent, $P(t)=1,$ the spin 
state is pure and it is located on the Bloch sphere.}

{The characteristic size of relatively high-density domains of the BEC is given by
the normalized participation ratio $\zeta(t)$: 
\begin{equation}
\zeta(t)\equiv\frac{1}{3}\left[\int_{-\infty}^{\infty }\left\vert {\bm \psi}(\mathbf{x}) \right\vert^{4}dx\right]^{-1}.
\label{eq:part-ratio}
\end{equation}%
The prefactor $1/3$ is chosen for consistency with the BEC 
width $w(t)=\left[N_{1}(t)w_{1}^{2}(t)+N_{2}(t)w_{2}^{2}(t)\right]^{1/2}.$ The latter characterizes its total spread, with:
\begin{eqnarray}
&&w_{\nu}^{2}(t)\equiv\int_{-\infty}^{\infty} \frac{x^{2}\left|\psi_{\nu}^{2}(\mathbf{x})\right|}{N_{\nu}(t)}dx,
\label{wnu} \\
&&N_{\nu}(t)\equiv\int_{-\infty}^{\infty}\left|\psi_{\nu}^{2}(\mathbf{x})\right|dx.
\nonumber
\end{eqnarray}
}
Here $w_{\nu}(t)$ is the component width, and $N_{\nu}(t)$ is the corresponding 
fraction of atoms with $N_{1}(t)+N_{2}(t)=1.$

\begin{figure}[t!]
\begin{center}
\includegraphics[width=0.6\columnwidth]{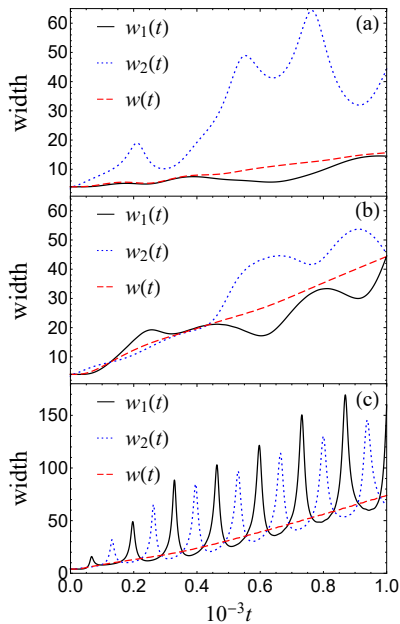}
\end{center}
\caption{{Width of the free space ($U(x)\equiv0$) BEC for $\Delta=0.01$ (a), $0.02$ (b), and $0.05$ (c).}
Here and below we use for numerical simulations $g=0.5.$}
\label{Fig:spin-up-w}
\end{figure}
\begin{figure}[]
\begin{center}
\includegraphics[width=0.6\columnwidth]{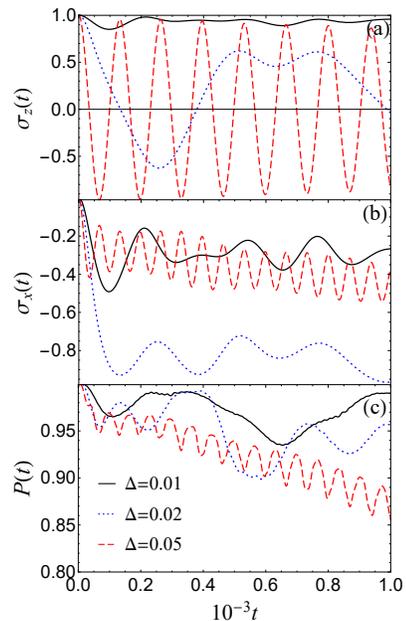}\\
\end{center}
\caption{{Evolution of the spin components along $z-$ (a) and $x-$ (b) axes, and of $P(t)=\sum_{i}\sigma_{i}^{2}(t)$ (c), 
for a free space BEC at $\Delta=0.01,0.02,0.05$}.}
\label{Fig:S-disor2}
\end{figure}

\section{Soliton evolution}

\subsection{Free space spin rotation and broadening} 

Here we present the analysis, which can be obtained also by summarizing the results known in 
the general theory of Rabi solitons in different systems, by expressing them in terms 
of the pseudo-spin $1/2$ BEC.
To focus on the most fundamental effects, we consider first a realization maximally different 
from the SU(2) symmetric Manakov-like case \cite{Manakov}, 
assuming $\tilde{g}=0$ {(the role of this cross-coupling will be discussed later on)}. 
The Zeeman field couples the spinor components and leads to
evolution of $\sigma_{i}(t)$ defined by Eq. (\ref{eq:sigmait}). This spin rotation causes a population redistribution between
components of the BEC spinor and, therefore, modifies its self-interaction energy. 
As a result, the Zeeman coupling 
and self-interaction energies become mutually related and shape of the soliton changes accordingly.
In order to better understand this process and for the qualitative analysis, we begin with the free motion
where $U(x)\equiv 0.$

Since the effect of the Zeeman field depends on the initial spin configuration,
for definiteness and simplicity, {here we consider an} initial state with $\sigma_{z}(0)=1.$ 
At $\Delta=0,$ {a stationary solution of Eq. (\ref{eq:GPE})} is
\begin{equation}
\psi_{1}({\mathbf x})=e^{-i\mu t}\frac{\sech(x/w_{0})}{\sqrt{2w_{0}}}, \quad \psi_{2}({\mathbf x})=0,
\label{eq:psi_equiv}
\end{equation}
where $w_{0}=2/g$ determines the energy scale \cite{zetavsw}, fixing the value of 
the chemical potential {to} $\mu=-g^{2}/8.$ {We indicate the} relevant {timescale as} $T_{\mu}{\equiv} 1/|\mu|,$
similar to the expansion time of a noninteracting wavepacket of the $w_{0}$ width. 
 
\begin{figure}[]
\begin{center}
\includegraphics[width=0.6\columnwidth]{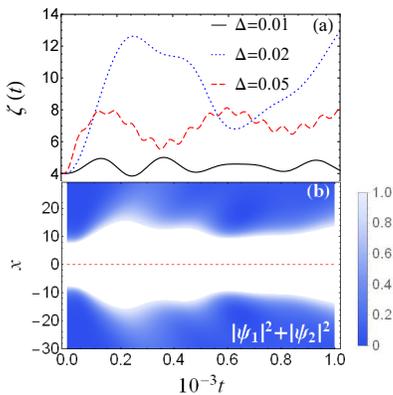}\\
\end{center}
\caption{
(a) {Participation ratio} of free space BEC defined by {Eq.} (\ref{eq:part-ratio}) for {$\Delta=0.01,0.02,$ and 0.05.}
{The broadening} of the soliton with the reorientation of the spin {is evident} for $\Delta$
{close to} the crossover value {$\Delta_{\rm cr}=|\mu|$}.
{(b) Density plot of the condensate in $(t,x)$ space, for $\Delta=0.02$.}
}
\label{Fig:spin-up}
\end{figure}

At nonzero $\Delta,$ the energy scale $\Delta$ comes into play, 
along with the corresponding spin rotation time $T_{\Delta }=2\pi/\Delta.$ 
Then, the competition between the Zeeman field and nonlinearity determines 
three possible regimes, namely for $\Delta$ smaller, larger, or of the order 
of the crossover value $\Delta_{\rm cr}\equiv|\mu|$.
Typical evolution patterns of the BEC parameters for the three regimes are shown respectively in  
Figs. \ref{Fig:spin-up-w}, \ref{Fig:S-disor2}, and \ref{Fig:spin-up}, and will be discussed in the following.

1. Weak Zeeman field, $\Delta \ll |\mu|,\,T_{\Delta}\gg T_{\mu}$. This regime is characterized by 
clearly different dynamics of the two spinor components (see Fig. \ref{Fig:spin-up-w}(a)), small 
amplitude spin rotations, $|\sigma_{x}(t)|,|\sigma_{y}(t)|\ll 1$ (Fig. \ref{Fig:S-disor2}), and 
a relatively small broadening of the wavepacket on the $T_{\Delta}$ time.  
{The latter is due to the fact that weak Zeeman fields only produce a small population 
of the second component, $N_{2}(t)\ll 1$, so that -- though this component spreads rapidly 
(with speed of the order of $g$, essentially due to the momentum-position uncertainty) -- 
it only produces a moderate increase of the total width $w$ (recall that the initial state $\psi_{1}(x)$ is stationary). 
Regarding the behavior of the spin components shown in Fig. \ref{Fig:S-disor2},} iterative 
solution of Eq. (\ref{eq:GPE}) corroborated by numerical results 
shows that at the initial stage, {$t\ll T_{\mu}$}, $\sigma_{x}(t)$ behaves as $2\Delta t^{2}\mu/3$ 
and the minimum value of $\sigma_{x}(t)$ is of the
order of $\Delta/\mu$, while the maximum value of $1-\sigma_{z}(t)$ is of the
order of $(\Delta/\mu)^{2}\ll\max(|\sigma_{x}(t)|)$.

2. {Crossover regime, $\Delta\sim |\mu|,\,T_{\Delta}\sim T_{\mu}$.} The Zeeman field becomes sufficiently strong to
rotate the spin by producing a sizable population $N_{2}(t).$ Consequently, the broadening of this component
decreases due to the self-attraction.  
In this case, the wavepacket broadening during a spin rotation period {$T_{\Delta}$} is of the
order of $w_{0}$, {and spin components which can trigger a 
substantial population exchange due to sufficient Zeeman energy changes, 
namely $\sigma_{z}(t)<0$ and $\sigma_{x}(t)\approx-1$,} can be achieved. 
Here, both components feature similar broadening with time while the dynamics of all relevant
quantities is rather irregular. {Numerical results show that 
although at $t>T_{\Delta}$ the initial soliton shape is already destroyed, its spin state
remains almost pure (see Fig. \ref{Fig:S-disor2}) with $P(t)\approx 1.$ Therefore,
$\psi_{1}(\mathbf{x})$ and $\psi_{2}(\mathbf{x})$ still remain approximately linearly  
dependent and the densities $|\psi_{1}(\mathbf{x})|^{2}$ and $|\psi_{2}(\mathbf{x})|^{2}$ 
have similar profiles.} 

3. {Strong magnetic field, $\Delta \gg |\mu|,\,T_{\Delta}\ll T_{\mu}$. In this case the two
spin components show regular oscillations,} with $\sigma_{z}(t)\approx\cos(\Delta t)$. 
The soliton width increases {almost linearly, each component being characterized by alternating 
periodic kicks (see Fig. \ref{Fig:spin-up-w}(c)). These kicks are due to the fact that 
each component spreads rapidly when its population $N_{\nu}(t)$ is minimal 
(see the discussion in the previous point 1), and then, when $N_{\nu}(t)\approx1$,} the spread rate 
decreases {significantly} due to the self-attraction. The analysis of the energy conservation yields 
that at a quarter of the Zeeman period, $t=T_{\Delta}/4,$ where $|N_{2}(t)-N_{1}(t)|\ll 1$, one obtains
$\sigma_x(\pi/2\Delta)\approx 2\mu/3\Delta,$ corresponding to the Zeeman energy required for 
this rotation (cf. Fig. \ref{Fig:S-disor2}(b)).

\subsection{Displacement driven by spin reorientation and disorder} 

A smooth disorder, like in the Lifshitz model \cite{Shklovskii}, 
is produced  at a long interval $L$ by a distribution of ${\cal N}\gg 1$ ``impurities'' with uncorrelated random
positions $x_{j}$ and mean linear density $\bar{n}={\cal N}/L$ as
\begin{equation}
U(x)=U_{0}\sum_{j=1}^{j=\cal N} s_{j}u(x-x_{j}).
\label{randomrealization}
\end{equation}%
Here $s_{j}=\pm 1$ is a random function of $j$ with 
{mean values} $\langle s_{j}\rangle=0$, {so that} $\langle U(x)\rangle =0$.
{Here we model the impurities as} $u(y)=\exp(-y^{2}/\xi^{2}),$ where $\xi$ is the corresponding width
(the results discussed in the following do not depend qualitatively on this specific choice). 
The motion of the BEC center of mass $X(t)$ (see Eq. (\ref{eq:Xt}))
is described by the Ehrenfest theorem \cite{Ehrenfest} as:
\begin{equation}
\frac{d^{2}X(t)}{dt^{2}}=F[{\bm\psi}]\equiv
-{\rm tr}\int_{-\infty}^{\infty} {\bm\rho}({\mathbf x}) U^{\prime}(x)dx,
\label{eq:Ehrensfest}
\end{equation}
where $F[{\bm\psi}]$ is the state-dependent force. 
For random $U(x),$ we choose as initial condition a stationary 
solution of Eq. (\ref{eq:GPE}) 
${\bm\psi}^{[d]}({\mathbf x}_{0})={[}\psi_{1}^{[d]}({\mathbf x}_{0}),0{]}^{\rm T}$ with $F[{\bm \psi}^{[d]}({\mathbf x}_{0})]=0,$
where ${\mathbf x}_{0}\equiv (x,0),$ corresponding to $\sigma_{z}(0)=1$ {as in the discussion of the free space case}.

{The disorder introduces a new energy-dependent timescale of elastic momentum relaxation related to particle backscattering
in a random potential. For a wavepacket, this timescale, being associated with  
the packet width in the momentum space, determines the time of free broadening of the packet till 
the localization effect will become essential.  In the Born scattering 
approximation this timescale is $\tau_{d}\equiv g/U_{0}^{2}\bar{n}\xi^{2},$  and the corresponding expansion 
length becomes $\ell = g\tau_{d}$ \cite{Gorkov,Richard}.}
We assume that the potential is weak such that $\ell\gg\,1/g,$ that is the initial width 
corresponding to $\psi_{1}^{[d]}({\mathbf x}_{0})$ [see Eq. (\ref{eq:part-ratio})], 
$\zeta(0)\approx\,w_{0}$, is due to the self-interaction rather than due to the conventional Anderson localization. 
In the following we shall consider relatively weak self-interactions, 
$g\xi\alt\,1$, to study wavepackets extended over several correlation lengths
of $U(x),$ where the effect of disorder is expected to be essential. 
{Notice that in this regime the potential is not able} to localize the condensate near a 
single minimum of $U(x)$, that is $\xi^{2}\lan U^{2}\ran^{1/2}\ll 1,$
where $\lan U^{2}\ran = U_{0}^{2}\bar{n}\xi \sqrt{\pi}.$
Also, we assume that $\min(\Delta,|\mu|)\tau_{d}\agt\,1$
hence the disorder does not influence strongly the short-term expansion.

In the following we will develop a simple scaling theory, describing this process qualitatively and then compare it
{with numerical} results. 
For broad states {as considered here}, the force $f_j$ imposed on the condensate by a single impurity
located at the point $x_{j}$ is given by:
\begin{equation}
f_{j}=\sqrt{\pi}U_{0}s_{j}\xi \left.\partial_{x}|{\bm\psi}(\mathbf{x})|^{2}\right|_{x=x_{j}}.
\label{eq:fimp}
\end{equation}
Disorder averaging, $\langle F^{2}[{\bm\psi}] \rangle \equiv \langle(\sum_{j}f_{j})^{2} \rangle${,} 
for the entire BEC yields \cite{Shklovskii} (see Appendix for details):
\begin{equation}
\langle F^{2}[{\bm\psi}] \rangle = \pi U_{0}^{2}\xi^{2}\bar{n}
\int_{-\infty}^{\infty} \left(\partial_{x}|{\bm\psi}(\mathbf{x})|^{2}\right)^{2} dx.
\label{eq:F2av}
\end{equation}

Equation (\ref{eq:F2av}) cannot be directly applied 
to the system {considered here} since
the {specific} initial equilibrium {condition} $F[{\bm\psi}^{[d]}({\mathbf x}_{0})]=0$ {is not} 
a subject of direct disorder-averaging. {Then, we  proceed as follows.}
At the initial stage of expansion ($t\ll {T_{\Delta}}$) of this strongly asymmetric set of the components we have: 
\begin{equation}
\psi_{1}^{[d]}(\mathbf{x})=\psi_{1}^{[d]}({\mathbf x}_{0}) + \delta\psi_{1}^{[d]}(\mathbf{x}), 
\quad 
\psi_{2}^{[d]}(\mathbf{x})= \delta\psi_{2}^{[d]}(\mathbf{x}),
\end{equation}
and the corresponding net force $\delta F$ acting on the condensate due to the $\delta\psi_{1}^{[d]}(\mathbf{x})-$term, 
is expressed as
\begin{equation}
\delta F = 
- 2{\rm Re}\int_{-\infty}^{\infty}\psi_{1}^{[d]}({\mathbf x}_{0})\delta\psi_{1}^{[d]}(\mathbf{x})U^{\prime}(x)dx.
\end{equation}
For {a} qualitative analysis, we can use a model of expansion of $\psi_{1}^{[d]}(\mathbf{x})$ by
assuming that the change in its
shape is solely due to change in the width $\delta w.$
With the same approach to the averaging of $\delta F$, we obtain (details are presented in the Appendix)
\begin{equation}
\langle(\delta F)^{2}\rangle = \frac{7\pi^{2}}{90} \frac{(\delta w)^2}{w_{0}^2}\langle F^{2}[{\bm\psi}]\rangle.
\label{eq:deltaF2av}
\end{equation}
{Here $\langle F^{2}[{\bm\psi}] \rangle = {\pi}U_{0}^{2}\xi^{2}\bar{n}g^{3}/30$ 
is the result of Eq. (\ref{eq:F2av})  for the state in Eq. (\ref{eq:psi_equiv}). It is
applicable for a weak disorder considered here,
where equilibrium shape $\psi_{1}^{[d]}$ is close to $\psi_{1}({\mathbf x})$ in Eq. (\ref{eq:psi_equiv}).}
Thus, the broadening of the wavepacket caused by switching on the Zeeman field results in
covering a different random potential and triggers its motion.

Now the three regimes of the spin evolution and broadening due to
the Zeeman field $\Delta\sigma_{x}/2$ leading to qualitatively similar regimes of its motion
in the random field, can be identified. {The main
feature} of the driven motion {is that} the force $\delta F$ 
needs a certain time to develop and then, it drives displacement of the condensate, $X(t)-X(0).$
For $\Delta\ll |\mu|,$ we have $|F[\psi_{2}^{[d]}(\mathbf{x})]|\ll |F[\psi_{1}^{[d]}(\mathbf{x})]|,$ 
the driven variations
in the density are weak and the position shows only small irregular oscillations.
At $\Delta\agt|\mu|$ (crossover and strong Zeeman couplings) {the contributions of }
$F[\psi_{2}^{[d]}(\mathbf{x})]$ and $F[\psi_{1}^{[d]}(\mathbf{x})]$ 
 are of the same order of magnitude, scaling  {as} 
$\left(U_{0}^{2}\xi^{2}\bar{n}g^{3}\right)^{1/2}.$
Therefore, {in the crossover regime,} the displacement during one Zeeman period can be estimated 
{from Eq. (\ref{eq:Ehrensfest})} as $\lan F^{2}\ran^{1/2} T_{\Delta}^{2},$ that is:
\begin{equation}
\sqrt{\langle X^{2}\left(T_{\Delta}\right)\rangle}
\sim U_{0}\xi\sqrt{\bar{n}}g^{-5/2}.
\end{equation}
{The} condition {for} a large displacement
during $T_{\Delta}$ triggering a long-distance propagation of the condensate
{corresponds to} $\sqrt{\langle X^{2}\left(T_{\Delta}\right)\rangle}\sim\,w_{0},$
that is $U_{0}\xi\sqrt{\bar{n}}g^{-3/2}\agt 1.$ 
The subsequent motion is a manifestation of {the} spin-position coupling due to {the} 
non-Manakov self-interaction in the BEC that can appear without SOC.

\begin{figure}[!t]
\begin{center}
\includegraphics[width=0.6\columnwidth]{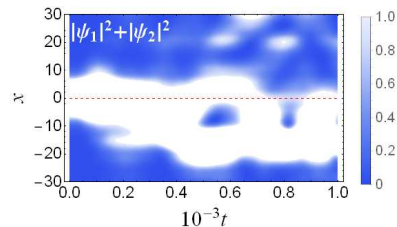}
\end{center}
\caption{
{Density plot of the condensate in a random potential for $\Delta=0.02$.
The initial values of the center-of-mass position and of the condensate width are $X(0)\simeq 0.19$ and  $\zeta(0)\simeq 3.25$, respectively.}
Note that for $t<100$ {($t<\tau_{d}$)} this plot is very similar to {that in} Fig. \ref{Fig:spin-up}{(b)}.}
\label{Fig:disor1}
\end{figure}

\begin{figure}[h!]
\begin{center}
\includegraphics[width=0.6\columnwidth]{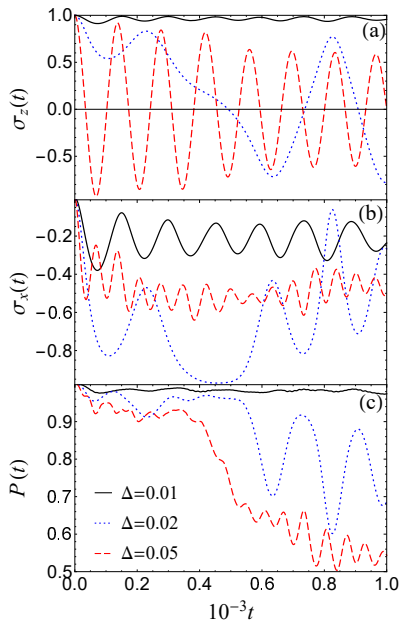}\\
\end{center}
\caption{{Evolution of the spin components along $x-$ (a) and $x-$ (b) axes, 
and of $P(t)$ (c), for $\Delta=0.01,0.02,$ and 0.05 (solid, dotted, and dashed lines, respectively).}
}
\label{Fig:S-disor2-2}
\end{figure}

For numerical calculations, in the following we fix $\xi=1$, $\bar{n} = 10/\xi$, and $U_{0}=0.01.$ 
With this choice,
the ground state of the condensate {extends over} several disorder correlation length, 
that is $2/g\gg\xi$ {(we recall that $g\equiv0.5$)}. 
Figure \ref{Fig:disor1} shows a {typical evolution of the total density} 
and demonstrates the net displacement and {the} change in the shape
of the condensate, including its possible splitting among two
potential minima. 
%
The spin evolution as presented in Fig. \ref{Fig:S-disor2-2} shows that the purity of the spin
state $P(t)$ is rapidly destroyed by the random potential due to the fact that $\psi_{1}^{[d]}({\mathbf x})$ and 
$\psi_{2}^{[d]}({\mathbf x})$ are linearly independent. 

\begin{figure}[t]
\begin{center}
\includegraphics[width=0.6\columnwidth]{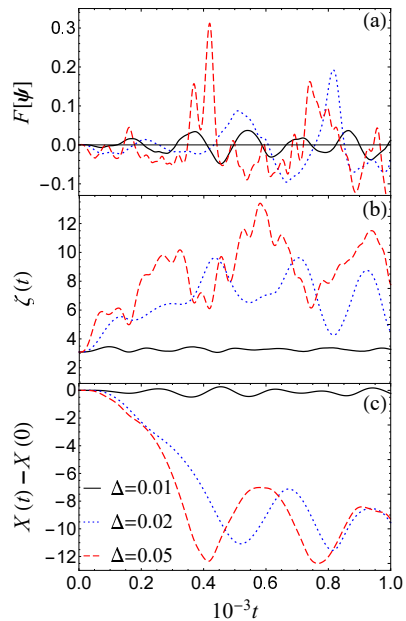}\\
\end{center}
\caption{Evolution of (a) the state-dependent force defined by Eq. (\ref{eq:F2av}),
(b) the {participation ratio,
(c)} the displacement of the center of mass $X(t)-X(0)$, for different values of $\Delta$.}
\label{Fig:disor2}
\end{figure}

The evolution of the force acting on the wavepacket, {its} size, and position are presented in Fig. \ref{Fig:disor2}.
{This figure} show that in a weak Zeeman field 
the condensate displacement is much smaller than its width,
the forces are weak, and the change in the width is small. Thus, the condensate shows
only small irregular oscillations near $X(0)$, as expected.
Figure \ref{Fig:disor2}(b) clearly demonstrates that also in the presence of disorder
broadening of the soliton depends on the Zeeman field.
The forces presented in Fig. \ref{Fig:disor2}(a) {have a clear correlation with the quantities shown in Figs.}
\ref{Fig:disor2}(b) and \ref{Fig:disor2}(c). Indeed, the force is large
when $\zeta$ is small and $d^{2}X/dt^{2}$ is large at large $F$. {In addition, a
comparison with the multipeak density profile in Fig. \ref{Fig:disor1}, 
confirms that the force is determined by $\zeta(t)$ in Eq. (\ref{eq:part-ratio}) rather than 
by total spread $w(t)>\zeta(t)$ in Eq. (\ref{wnu}).}
Although the random motion considerably
depends on the realization of $U(x),$ this dependence is only quantitative,
and the entire qualitative analysis remains valid independent of the given realization.

\begin{figure}[t]
\begin{center}
\includegraphics[width=0.6\columnwidth]{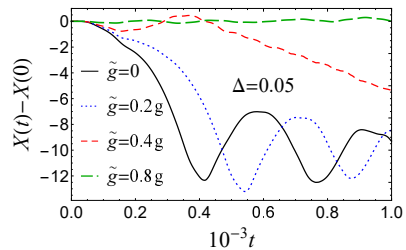}\\
\end{center}
\caption{Time-dependent displacement of the center of mass 
for different values of $\tilde{g}$ and $\Delta=0.05.$
}
\label{Fig:gtilde}
\end{figure}

Having discussed a realization with {$\tilde{g}=0$ and single-component initial conditions,} 
we proceed with a brief analysis of other {possible scenarios}. We begin
with the same initial condition and different $\tilde{g},$ as presented 
in Fig. \ref{Fig:gtilde}, demonstrating that  
with the increase in $\tilde{g},$
the driving effect of the Zeeman field decreases, 
and vanishes for the SU(2) symmetry  \cite{Tokatly2010,Tokatly2013,Amico2000}, where
$\tilde{g}=g.$ In this limit, the spin rotation does not require energy to 
modify the self-interaction since the condensate rotates without change in its 
shape as ${\bm\psi}^{[d]}(\mathbf{x})=\psi_{1}^{[d]}({\mathbf x}_{0})\left(\cos(\Delta t/2),\sin(\Delta t/2)\right)^{\rm T},$ 
and no net force $\delta F$ appears as a result. 
As far as the role of the initial conditions {is concerned, we notice that there is an} infinite number
of states ${\bm\psi}^{[d]}({\mathbf x}_{0})=\left[\psi_{1}^{[d]}({\mathbf x}_{0}),\psi_{2}^{[d]}({\mathbf x}_{0})\right]^{\rm T}$
satisfying the stationarity condition $F[{\bm\psi}^{[d]}({\mathbf x}_{0})]=0.$ When {a 
Zeeman field is applied along the $x-$axis, a} precession around this axis begins, 
{which in turn modifies the density distribution, then} leading to a nonzero force and causing further dynamics. 
For linearly independent $\psi_{1}^{[d]}({\mathbf x}_{0})$ and $\psi_{2}^{[d]}({\mathbf x}_{0}),$ the spin rotation 
leads to {a} change in the self-interaction energy, and {in general a} net force appears
for the SU(2) coupling{, as well}. This guarantees that the triggering of the motion 
of initially stationary BEC by a Zeeman field as discussed in the present paper is in fact a general feature of 
self-interacting pseudo-spin $1/2$ condensates.

\section{Conclusions} 

We have demonstrated that the motion of the center-of-mass of a self-attractive spinor
Bose-Einstein condensate can be caused by {the} joint effect of the spin precession in a Rabi-like Zeeman field
and {the presence of} an external potential considered here in a random form as an example. 
The broadening of the condensate caused by the spin rotation leads
to a net force acting on it and triggers its motion. Thus, the spin evolution can drive changes in the condensate
position even in the absence {of spin-orbit} coupling. These results hint at possible interesting extensions 
of the present study, including the theory of multidimensional and multisoliton settings.

\section*{Acknowledgments}
We acknowledge support by the Spanish Ministry of Economy, Industry and 
Competitiveness (MINECO) and the European Regional Development Fund FEDER 
through Grant No. FIS2015-67161-P (MINECO/FEDER, UE) and the 
Basque Government through Grant No. IT986-16. S.M. was partially supported by the Swiss National 
Foundation SCOPES Project No. IZ74Z0-160527. The work of B.A.M. was supported, in part, by the Binational (US-Israel)
Science Foundation through project No. 2015616, and by the
Israel Science Foundation through Grant No. 1286/17.
We are grateful to V.V. Konotop and Y. V. Kartashov for discussions and valuable comments.

\appendix

\renewcommand\thefigure{\thesection.\arabic{figure}}    

\section{Disorder averaging}

\setcounter{figure}{0} 

Here we describe the disorder-averaging calculation of the force
acting on the condensate in a random field. For definiteness, we will omit
the time dependence and consider only relevant coordinate-dependences using
the same notations as in the main text. 

\begin{figure}[tbp]
\begin{center}
\includegraphics[width=0.85\columnwidth]{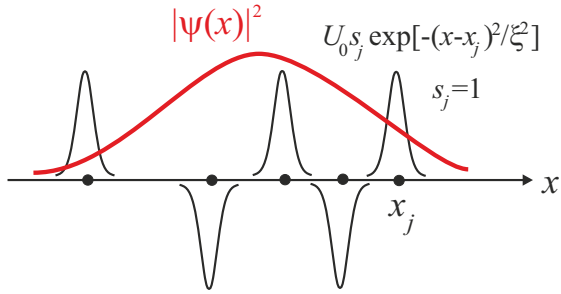}
\end{center}
\caption{Schematic plot of the BEC density $|{\bm\psi }(x)|^{2}$ and impurity potential. Positions of impurities 
are marked with filled circles.}
\label{disorder}
\end{figure}

We consider a random potential produced by distribution of "impurities"
with "white-noise" uncorrelated  random positions $x_{j}$ and mean
linear density $\bar{n}$ of the form 
\begin{equation}
U(x)=U_{0}\sum_{j}s_{j}u(x-x_{j}),  \label{randomrealization}
\end{equation}%
where $s_{j}=\pm 1$ is a random function of $j$ with mean value $\langle
s_{j}\rangle =0$, so that $\langle U(x)\rangle =0,$ and Gaussian shape of $u(x-x_{j})=\exp
(-(x-x_{j})^{2}/\xi ^{2}),$ where $\xi $ is the corresponding width. We
begin with the effect of a single impurity located at the point $x_{j}$ on the 
condensate energy and applied force for broad states of our interest, see Fig. \ref{disorder}.
The ``single-impurity'' interaction energy $v_{j}$ and force $f_{j}$ are given by: 
\begin{eqnarray}
v_{j}&=&U_{0}s_{j}\int_{-\infty }^{\infty }u(x-x_{j})|{\bm\psi }({x})|^{2}dx,  
\label{Aeq:Ehrensfest} \\
f_{j}&=&-U_{0}s_{j}\int_{-\infty }^{\infty }|{\bm\psi }({x})|^{2}u^{\prime }(x-x_{j})dx.
\end{eqnarray}%
For the chosen Gaussian impurity shape we obtain
\begin{equation}
v_{j}=\sqrt{\pi }U_{0}s_{j}\xi |{\bm\psi }(x_{j})|^{2}.
\end{equation}%
For the force we expand the density in the vicinity of the $x_{j}$ point as $|%
{\bm\psi }({x})|^{2}=|{\bm\psi }(x_{j})|^{2}+\left.
\left(d|{\bm\psi }({x})|^{2}/dx\right)\right\vert _{x=x_{j}}\left(
x-x_{j}\right) $ and obtain

\begin{equation}
f_{j}=\sqrt{\pi }U_{0}s_{j}\xi \left. 
\frac{d}{dx}|{\bm\psi }({x})|^{2}\right\vert _{x=x_{j}}.
\end{equation}

To produce the disorder averaging for the uncorrelated distribution of impurities, $\langle
F^{2}[{\bm\psi}]\rangle\equiv\langle(\sum_{j}f_{j})^{2}\rangle$ and $\langle
V^{2}[{\bm\psi}]\rangle\equiv\langle(\sum_{j}v_{j})^{2}\rangle,$ for the entire
condensate, we use the technique presented in detail in Ref. \cite{Shklovskii}.
With this approach the sum over impurities for a function $\chi(x)$, $p_{\chi}\equiv \sum_{j}\chi(x_{j})$ 
is presented as an integral, in our case in the form:%
\begin{equation}
\langle p_{\chi}^{2} \rangle =\bar{n}\int 
\delta (x-x^{\prime })\chi(x)\chi(x^{\prime })dxdx^{\prime }.
\end{equation}%
Thus, we arrive at the transformation:%
\begin{equation}
\langle p_{\chi}^{2} \rangle =\bar{n}\int \chi^{2}(x)dx,
\end{equation}%
and obtain for the energy and the force:
\begin{eqnarray}
\langle V^{2}[{\bm\psi }]\rangle  &=&\pi U_{0}^{2}\xi ^{2}\bar{n}%
\int_{-\infty }^{\infty }|{\bm\psi }({x})|^{4}dx \label{Aeq:F2av} \\
&=&\frac{\pi }{3}U_{0}^{2}\xi ^{2}\bar{n}\frac{1}{\zeta }, \nonumber \\   
\langle F^{2}[{\bm\psi }]\rangle  &=&\pi U_{0}^{2}\xi ^{2}\bar{n}%
\int_{-\infty }^{\infty }
\left(\frac{d}{dx}|{\bm\psi }({x})|^{2}\right) ^{2}dx,
\end{eqnarray}%
where the participation ratio $\zeta $ is defined by Eq. (\ref{eq:part-ratio}) of the main text. 
For the wavefunction in Eq. (\ref{eq:psi_equiv}) of the main text:  
\begin{equation}
\psi _{1}({{x}})=\frac{\sech(x/w_{0})}{\sqrt{2w_{0}}},\quad \psi _{2}(%
{{x}})=0,  \label{Aeq:psi_equiv}
\end{equation}%
these equations yield: 
\begin{eqnarray}
\langle V^{2}[{\bm\psi }]\rangle  &=&\frac{\pi }{6}U_{0}^{2}\xi ^{2}\bar{n}g,
\label{V2} \\
\langle F^{2}[{\bm\psi }]\rangle  &=&\frac{\pi }{30}U_{0}^{2}\xi ^{2}\bar{n}%
g^{3}.
\label{F2}
\end{eqnarray}%
Note that these relations can readily be understood by using the
basic fluctuations theory for non-correlated ensembles. For this purpose we 
recall that the relevant spatial scale of the BEC density distribution is $\zeta.$ Then, for a 
qualitative scaling analysis, the fluctuations in $V[{\bm\psi}]$ and $F[{\bm\psi}]$ 
can be presented in terms of the difference in the number of
impurities with $s_{j}=1$ and $s_{j}=-1$ at this spatial scale. The
fluctuation of the square of this difference, relevant for $\langle V^{2}[{\bm\psi }]\rangle$  
and $\langle F^{2}[{\bm\psi }]\rangle,$ is of the order $\bar{n}\zeta,$
which yields $\langle V^{2}[{\bm\psi }]\rangle \sim U_{0}^{2}\xi ^{2}\bar{n}%
/\zeta $ and $\langle F^{2}[{\bm\psi }]\rangle \sim U_{0}^{2}\xi ^{2}\bar{n}%
/\zeta^{3},$  in agreement with Eqs. (\ref{V2}) and (\ref{F2}). 

The above disorder-averaging procedure of the force is not directly applicable near the
equilibrium since at $t=0$ a special condition $F[{\bm\psi }%
^{[d]}(x)]=0$ is satisfied. Thus, we have to consider
variation of the force $\delta F$ due to the variations of the BEC wavefunction in the form:%
\begin{equation}
\delta F=-2\mathrm{Re}\int_{-\infty }^{\infty }\psi _{1}^{[d]}({{x}}%
_{0})\delta \psi _{1}^{[d]}({x})U^{\prime }(x)dx.
\end{equation}%
Due to a change in the width $\delta w$ this variation for wavefunction in Eq. (\ref{eq:psi_equiv}) 
of the main text becomes: 
\begin{equation}
\delta \psi _{1}^{[d]}({x})=\delta w\frac{x\sinh (x/w_{0})}{\sqrt{2}%
w_{0}^{5/2}\cosh ^{2}(x/w_{0})},
\end{equation}%
and we arrive at Eq. (\ref{eq:deltaF2av}) of the main text:  
\begin{equation}
\langle (\delta F)^{2}\rangle =\frac{7\pi ^{2}}{90}\frac{(\delta w)^{2}}{%
w_{0}^{2}}\langle F^{2}[{\bm\psi }]\rangle,
\end{equation}
with $\langle F^{2}[{\bm\psi }]\rangle$ from Eq. (\ref{F2}).


\begin{thebibliography}{99}
\bibitem{zetavsw} {The factor 1/3 in Eq. (\ref{eq:part-ratio}) assures that for the soliton shape
in Eq. (\ref{eq:psi_equiv}) one obtains $\zeta(0)=w_{0}$.}
\bibitem{length} Note that for a zero-width potential with $\xi\rightarrow\,0$, the force vanishes even if
the mean value of $\lan U^{2}\ran$ is still finite. Therefore, in this limit $X(t)=X(0).$
\end{thebibliography}
\end{document}